\documentclass[letterpaper, 10 pt, conference]{ieeeconf}  

\IEEEoverridecommandlockouts                              

\usepackage{graphicx}
\usepackage{amsmath} 
\usepackage{amssymb}  
\usepackage{booktabs}
\usepackage{threeparttable}
\usepackage{multirow}
\usepackage{tabularx}
\usepackage{dsfont}
\usepackage{cite}
\usepackage{url}

\newcommand{\argmax}{\mathop{\rm arg~max}\limits}

\title{\LARGE \bf
Evaluating Classifier Confidence for Surface EMG Pattern Recognition
}

\author{Akira Furui$^{1}$
\thanks{This work was not supported by any organization.}
\thanks{$^{1}$Akira Furui is with the Graduate School of Advanced Science and Engineering, Hiroshima University, Higashihiroshima 739-8527, Japan
        (e-mail: {\tt\small akirafurui@hiroshima-u.ac.jp}).}%
}

\begin{document}

\maketitle
\thispagestyle{empty}
\pagestyle{empty}

\begin{abstract}
Surface electromyogram (EMG) can be employed as an interface signal for various devices and software via pattern recognition.
In EMG-based pattern recognition, the classifier should not only be accurate, but also output an appropriate confidence (i.e., probability of correctness) for its prediction.
If the confidence accurately reflects the likelihood of true correctness, then it will be useful in various application tasks, such as motion rejection and online adaptation.
The aim of this paper is to identify the types of classifiers that provide higher accuracy and better confidence in EMG pattern recognition.
We evaluate the performance of various discriminative and generative classifiers on four EMG datasets, both visually and quantitatively.
The analysis results show that while a discriminative classifier based on a deep neural network exhibits high accuracy, it outputs a confidence that differs from true probabilities.
By contrast, a scale mixture model-based classifier, which is a generative classifier that can account for uncertainty in EMG variance, exhibits superior performance in terms of both accuracy and confidence.
\end{abstract}

\section{Introduction}

Human-machine interfaces that utilize surface electromyogram (EMG) facilitate the intuitive operation of devices and application software.
EMG signals are electrical activities generated during muscle contraction, and various motion-specific signal patterns can be obtained by attaching electrodes to multiple locations on the skin surface.
These interfaces utilizing EMG signals have been applied in various fields, including myoelectric prosthetic hands, human-computer interactions, and rehabilitation engineering~\cite{Asghar2022-fu}.

To realize such interfaces, machine learning techniques, particularly those based on pattern recognition, are typically employed.
Specifically, mapping between the measured EMG patterns and class labels (e.g., executed motions) is learned using training data, and then class labels on test data are predicted.
Various classifiers have been used to achieve this previously, and they can be broadly categorized into discriminative and generative classifiers.
Each type presents its own advantages and disadvantages; however, the authors recently reported the effectiveness of a generative classifier that can account for uncertainty in EMG variance~\cite{Furui2021-ts}.

In EMG-based pattern recognition, classifiers must demonstrate not only high accuracy, but also confidence in their decisions.
For example, rejecting motions with low confidence can prevent unintended control~\cite{Scheme2013-mf}.
The use of high-confidence test data for sequential learning can result in effective online adaptation.
In general, prediction confidence is defined by the posterior probability of the class prediction output by the probabilistic classifier.
This probability ideally reflects the likelihood of the true correctness.

However, a recent study demonstrated that modern deep neural network-based classifiers tend to exhibit overconfidence compared with classical classifiers, despite their improved accuracy~\cite{Guo2017-ml}.
In EMG pattern recognition, researchers have suggested that whereas different classifiers may exhibit similar classification accuracies, their underlying confidence profiles may differ substantially~\cite{Scheme2015-tg}.
Therefore, identifying EMG classifiers that provide good confidence may facilitate the development of reliable interfaces.

In this paper, we examine the confidence of classifiers used in EMG pattern recognition across multiple datasets.
Specifically, we evaluate the accuracy and confidence of several discriminative and generative classifiers with different levels of complexity and characteristics.
The classifier confidence is analyzed visually and quantitatively.
By performing a series of analyses, we aim to clarify the classifier types that provide both high accuracy and good confidence estimates.

\section{Materials and Methods}

\subsection{Datasets}

We used four publicly available EMG datasets on upper-limb motions.
Table~\ref{tab:dataset} presents the characteristics of each dataset.
Datasets I, II, and III were obtained from \cite{Khushaba2012-zo}, \cite{Khushaba2013-mn}, and \cite{Khushaba2012-ii}, respectively.
Dataset IV was extracted from the \textit{putEMG}~\cite{Kaczmarek2019-ns} dataset.
In this dataset, the participants performed three distinct tasks, with the task referred to as \textit{repeats\_short}, as utilized in the present study.
Although three elastic bands, each with eight electrodes, were used for EMG measurements, only the middle band was used.

For feature extraction, each dataset was subjected full-wave rectification and smoothing using a second-order low-pass Butterworth filter with a cutoff frequency of 2 Hz. 
The classifiers were trained and tested individually for each participant. 
Each dataset involved multiple trials (motion repetitions), with the first two trials serving as the training set and the remaining trials the test set.

\begin{table*}[]
        \centering
        \caption{Dataset information}
        \begin{tabular}{lllllll}
        \toprule
        Dataset & \# Motions ($C$) & \# Electrodes ($D$) & Sampling frequency & \# Participants & Training trials & Test trials \\ \midrule
        I & 10 & 2 & 4,000 Hz & 10 & 1, 2 & 3, 4, 5, 6 \\
        II & 14 & 8 & 4,000 Hz & 8 & 1, 2 & 3, 4 \\
        III & 15 & 8 & 4,000 Hz & 8 & 1, 2 & 3 \\
        IV & 7 & 8 & 5,120 Hz & 44 & 1, 2 & 3, 4, 5, 6 \\ \bottomrule
        \label{tab:dataset}
        \end{tabular}
\end{table*}

\subsection{Classifiers}

Let $\mathbf{x} \in \mathbb{R}^D$ be an EMG feature pattern recorded from $D$ electrodes and $y \in \{1, \ldots, C\}$ be the corresponding class label (target motion).
In this experiment, three discriminative and three generative classifiers were trained and tested on each dataset to evaluate their performance and confidence.

\subsubsection{Discriminative Classifiers}

Discriminative classifiers model the posterior probability for each class (motion). 
In other words, they directly estimate the conditional probability $p(y|\mathbf{x})$ through training.

\begin{itemize}
        \item \textbf{Linear logistic regression (LLR)} is a linear probabilistic classification model and a type of generalized linear model that employs logits as the link function. The model has a unique solution and can be trained using optimization algorithms such as quasi-Newton methods. We applied a weight decay of 0.01.
        \item \textbf{Multilayer perceptron (MLP)} is a classifier composed of multiple layers of perceptrons and is the standard neural network model. In this experiment, we defined one hidden layer comprising 50 units and employed a rectifier linear unit (ReLU) as its nonlinear activation function. This model was trained using the Adam optimization algorithm with a learning rate of 0.001 and a batch size of 128. We applied the weight decay of 0.0001.
        \item \textbf{Deep multilayer perceptron (Deep MLP)}, which is defined as a deep network structure with multiple hidden layers, is expected to exhibit superior learning capabilities compared with MLP, as the nonlinear function is repeated several times. In this experiment, three hidden layers were established using 100, 50, and 25 units in each hidden layer. The nonlinear activation function and optimization algorithm were the same as those used for the MLP. In addition, batch normalization was applied to each hidden layer to stabilize the learning process. 
\end{itemize}

\subsubsection{Generative Classifiers}

Generative classifiers model the joint probability $p(\mathbf{x}, y) = p(\mathbf{x} | y) p(y)$ of the inputs and outputs.
Subsequently, the posterior probability for each class can be calculated using Bayes' rule as follows:
\begin{align}
        p(y | \mathbf{x}) = \frac{p(\mathbf{x} | y) p(y)}{p(\mathbf{x})}.
\end{align}
A classifier can be designed by establishing an appropriate distribution for the class-dependent generative model $p(\mathbf{x} | y )$.

\begin{itemize}
        \item \textbf{Linear discriminant analysis (LDA)}, which is a method applied not only for classification but also for a wide range of applications such as dimensionality reduction, can be interpreted as a generative classifier based on Gaussian distributions with a shared covariance matrix among classes. Therefore, the generative model of class $c$ in LDA is expressed as
        \begin{align}
                p(\mathbf{x} | y = c) = \mathcal{N}(\mathbf{x}|\boldsymbol{\mu}_c, \boldsymbol{\Sigma}),
        \end{align}
        where $\mathcal{N}(\mathbf{x}|\boldsymbol{\mu}, \boldsymbol{\Sigma})$ represents a Gaussian distribution with a mean $\boldsymbol{\mu} \in \mathbb{R}^D$ and covariance matrix $\boldsymbol{\Sigma} \in \mathbb{R}^{D \times D}$.
        This model is typically used in EMG-based systems owing to its simple structure and low computational cost.
        \item \textbf{Quadratic discriminant analysis (QDA)} is a generative classifier that uses a Gaussian distribution with a unique covariance matrix for each class: 
        \begin{align}
                p(\mathbf{x} | y = c) = \mathcal{N}(\mathbf{x}|\boldsymbol{\mu}_c, \boldsymbol{\Sigma}_c).
        \end{align}
        Unlike LDA, which can only learn linear decision boundaries, QDA can learn quadratic boundaries. QDA is equivalent to a Gaussian mixture model-based classifier with a single-mixture component. 
        \item \textbf{Scale mixture model-based classifier (SMMC)}~\cite{Furui2021-ts}, which is a non-Gaussian classifier, captures the characteristics of EMG generation more effectively. The classifier is composed of a linear combination of scale mixture models~\cite{Furui2019-ho}, which assume a hierarchical probability structure for the signal variance, and can account for the uncertainty in the EMG variance. The generative model for class $c$ in this classifier is expressed using the latent variable $u_{ck}$, which generates stochastic fluctuations in EMG variance:
        \begin{align}
                p(\mathbf{x} | y = c) 
                &= \sum_{k=1}^{K_c} \pi_{ck} \int \mathcal{N}(\mathbf{x}|\boldsymbol{\mu}_{ck}, u_{ck} \boldsymbol{\Sigma}_{ck}) \nonumber \\
                & \quad \quad \quad \quad \quad \times\mathrm{IG} \left(u_{ck} \middle| \frac{\nu_c}{2}, \frac{\nu_c}{2}\right) \mathrm{d} u_{ck},
        \end{align}
        where $K_c$ is the number of mixing components, $\pi_{ck}$ is the mixing coefficient, $\nu_c$ is the degree-of-freedom parameter, and $\mathrm{IG} (u | a, b)$ denotes the inverse gamma distribution with shape $a$ and scale $b$.
        The authors showed that SMMC can classify EMG patterns with higher accuracy than the existing classification models~\cite{Furui2021-ts}.
        Furthermore, the authors presented a technique for automatically determining the hyperparameters, $K_c$ and $\nu_c$. 
        In this paper, however, we fixed them as $K_c = 1$ and $\nu_c = 0.1$ for simplification.
\end{itemize}

\begin{figure*}[!t]
        \centering
        \includegraphics[width=1.0\hsize]{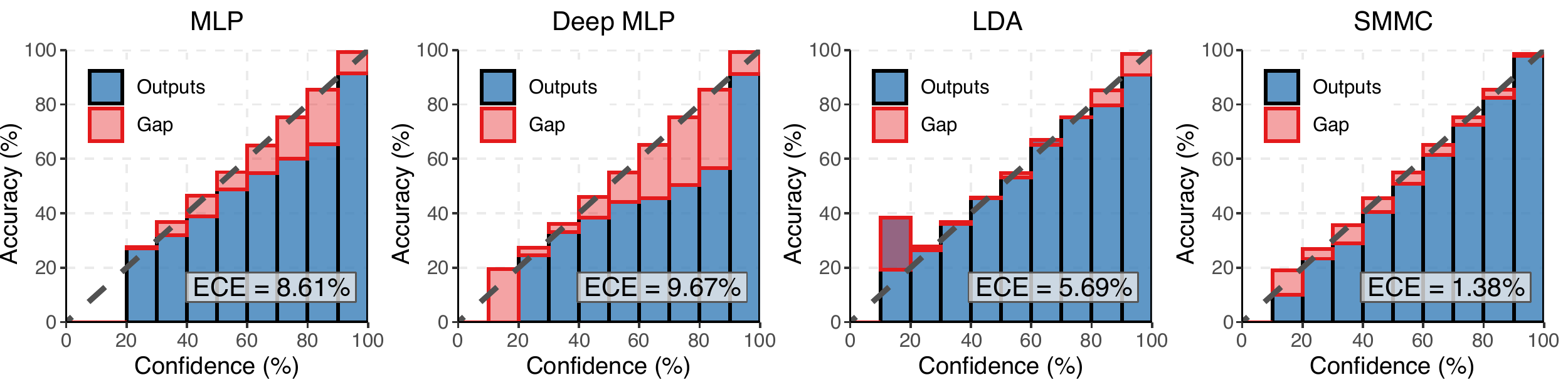}
        \caption{Reliability diagrams for MLP, Deep MLP, LDA, and SMMC on Dataset III. Red bar represents calibration gap for each bin.}
        \label{fig:rd}
\end{figure*}

\subsection{Evaluation}

After training each classifier on the training set, we calculated its accuracy using the test set.
In addition to the test accuracy, we evaluated the confidence of the class predictions and compared them across classifiers.

Given a test data point $\mathbf{x}_n$, a class prediction $\hat{y}_n$ and its associated confidence $\hat{p}_n$ can be obtained using the posterior probability, which is the output of the classifier, as follows:
\begin{align}
        \hat{y}_n &= \argmax_c p(y_n = c | \mathbf{x}_n),\\
        \hat{p}_n &= \max_c p(y_n = c | \mathbf{x}_n).
\end{align}
Ideally, the confidence estimate $\hat{p}_n$ should reflect the probability of the classification result being correct.
As an example, given 100 predictions, if the confidence of each prediction is 0.7, then 70 predictions should be classified correctly.
Thus, a classifier that provides adequate confidence is described as \textit{well calibrated}.
We evaluated the confidence of each classifier using the following methods.

\textbf{Reliability diagrams} are visual representations of the classifier calibration~\cite{Guo2017-ml}.
Given $N$ test data points, their prediction confidence is first segmented into $M$ interval bins, thereby 
Let $B_m$ be the set of indices corresponding to the data points in each bin $m \in \{1,\ldots,M\}$, then the average accuracy $\mathrm{acc}(B_m)$ and average confidence $\mathrm{conf}(B_m)$ for each $B_m$ are calculated as follows:
\begin{align}
        \mathrm{acc}(B_m) &= \frac{1}{|B_m|} \sum_{i \in B_m} \mathds{1} (\hat{y}_i = y_i), \\
        \mathrm{conf}(B_m) &= \frac{1}{|B_m|} \sum_{i \in B_m} \hat{p}_i,
\end{align}
where $y_i$ is the true class label, $\mathds{1}(\cdot)$ is the indicator function, and $|B_m|$ represents the size of $B_m$.
Finally, a plot is created with $\mathrm{conf}(B_m)$ along the horizontal axis and $\mathrm{acc}(B_m)$ along the vertical axis.
Thus, a perfectly calibrated classifier will exhibit $\mathrm{acc}(B_m) = \mathrm{conf}(B_m)$ for all $m$, and the plot will be diagonal.
Any deviation from the diagonal implies a miscalibration.
In this experiment, the number of bins was set to $M=10$.

\textbf{Expected calibration error (ECE)} is an approximation of the difference in expectations between confidence and accuracy~\cite{Naeini2015-pf}.
This metric can be calculated using the average of the bins' accuracy and confidence as follows:
\begin{align}
        \mathrm{ECE} = \sum_{m=1}^M \frac{|B_m|}{N} \left| \mathrm{acc}(B_m) - \mathrm{conf}(B_m) \right|.
\end{align}
The difference between the accuracy and confidence for a given bin represents a calibration gap, which corresponds to the deviation from the diagonal in reliability diagrams.

\textbf{Maximum calibration error (MCE)} is an approximation of the worst-case deviation between confidence and accuracy~\cite{Naeini2015-pf}.
Similar to ECE, this metric can be calculated as
\begin{align}
        \mathrm{MCE} = \max_{m \in \{1,\ldots,M\}} \left| \mathrm{acc}(B_m) - \mathrm{conf}(B_m) \right|.
\end{align}
The MCE can be regarded as the largest calibration gap across all bins in reliability diagrams.

\section{Results}

Fig.~\ref{fig:rd} shows the reliability diagrams for MLP, Deep MLP, LDA, and SMMC on Dataset III.
The dashed diagonal line in each diagram indicates perfect calibration.
The red bar represents the calibration gap for each bin, where the gap below and above the diagonal indicates overconfidence and underconfidence, respectively.
Additionally, the ECE metric corresponding to each reliability diagram is shown in the figure, which represents the weighted average of the gaps.

\begin{figure}[!t]
        \centering
        \includegraphics[width=0.85\hsize]{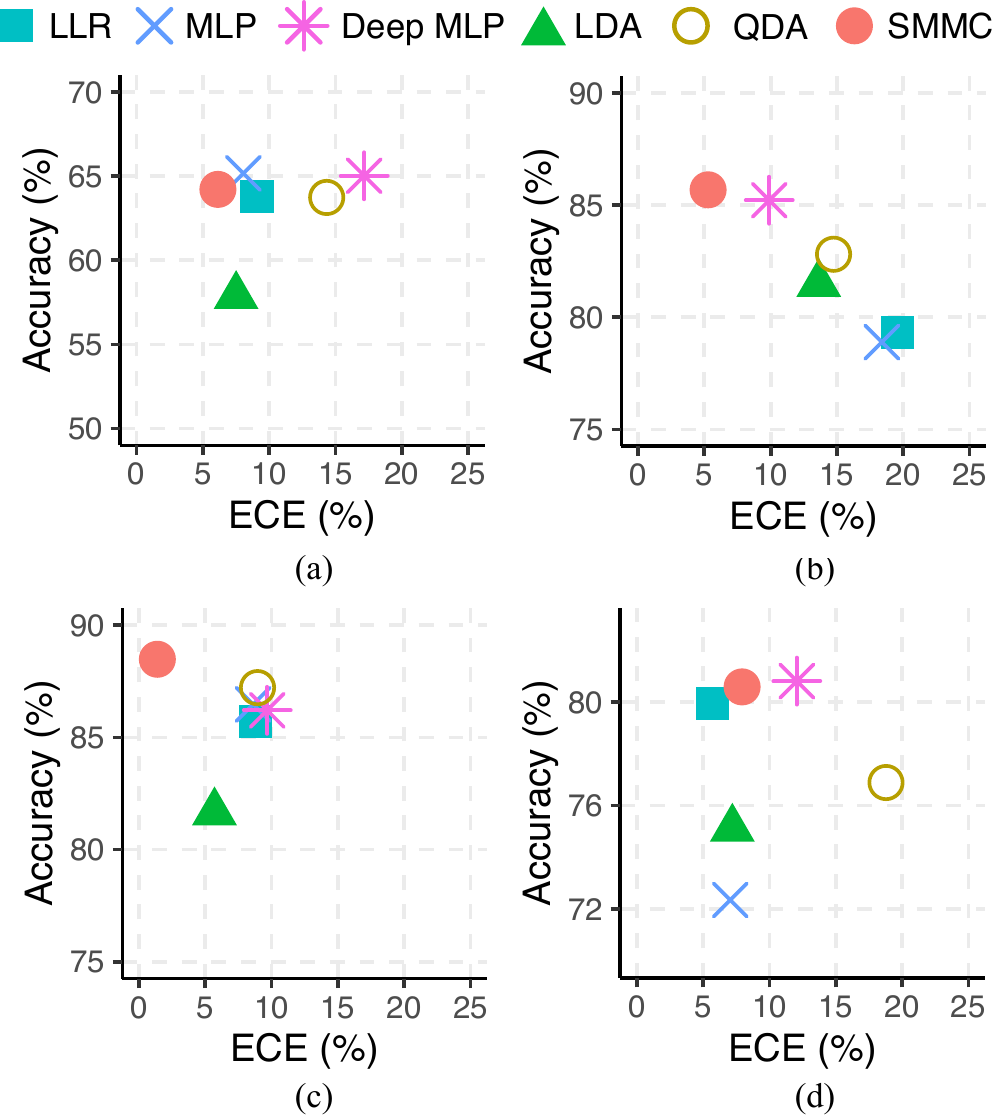}
        \caption{Accuracy vs. ECE plot for each dataset. (a) Dataset I. (b) Dataset II. (c) Dataset III. (d) Dataset IV.}
        \label{fig:acc_ece}
\end{figure}

Table~\ref{tab:metrics} lists the accuracy and confidence metrics for each classifier.
The best metric values for each dataset are highlighted in bold font.
Note that higher accuracy and lower ECE/MCE indicate better results.
The relationship between accuracy and ECE is shown in a scatterplot in Fig.~\ref{fig:acc_ece}.
Based on the property of each metric, the classifiers plotted in the upper-left corner of the scatterplot exhibit higher accuracy and are well-calibrated.

\begin{table}
        \centering
        \caption{Performance metrics on EMG datasets}
        \begin{tabular}[!t]{ccccc}
                \toprule
                Dataset & Classifier & Accuracy ($\uparrow$) & ECE ($\downarrow$) & MCE ($\downarrow$)\\
                \midrule
                 & LLR & 63.78\% & 9.09\% & 13.07\%\\
                
                 & MLP & \textbf{65.17\%} & 8.04\% & 16.35\%\\
                
                 & Deep MLP & 65.02\% & 17.16\% & 26.33\%\\ 
                
                 & LDA & 57.88\% & 7.50\% & \textbf{12.77}\%\\
                
                 & QDA & 63.74\% & 14.33\% & 20.67\%\\
                
                \multirow{-6}{*}{\centering\arraybackslash I} & SMMC & 64.21\% & \textbf{6.13\%} & 27.53\%\\
                \cmidrule{1-5}
                 & LLR & 79.32\% & 19.56\% & 42.16\%\\
                
                 & MLP & 78.89\% & 18.39\% & 32.34\%\\
                
                 & Deep MLP & 85.22\% & 9.89\% & 28.41\%\\ 
                
                 & LDA & 81.49\% & 13.63\% & 30.99\%\\
                
                 & QDA & 82.81\% & 14.74\% & 38.51\%\\
                
                \multirow{-6}{*}{\centering\arraybackslash II} & SMMC & \textbf{85.68\%} & \textbf{5.29\%} & \textbf{15.93\%}\\
                \cmidrule{1-5}
                 & LLR & 85.70\% & 8.78\% & 15.75\%\\
                
                 & MLP & 86.47\% & 8.61\% & 20.09\%\\
                
                 & Deep MLP & 86.21\% & 9.67\% & 28.83\%\\ 
                
                 & LDA & 81.66\% & 5.69\% & 19.06\%\\
                
                 & QDA & 87.22\% & 8.94\% & 25.45\%\\
                
                \multirow{-6}{*}{\centering\arraybackslash III} & SMMC & \textbf{88.48}\% &\textbf{1.38\%} & \textbf{8.99\%}\\
                \cmidrule{1-5}
                 & LLR & 79.95\% & \textbf{5.69\%} & 77.21\%\\
                
                 & MLP & 72.35\% & 7.04\% & 15.41\%\\
                
                 & Deep MLP & \textbf{80.81\%} & 12.06\% & 28.02\%\\ 
                
                 & LDA & 75.11\% & 7.20\% & \textbf{13.25\%}\\

                 & QDA & 76.89\% & 18.79\% & 37.42\%\\
                
                \multirow{-6}{*}{\centering\arraybackslash IV} & SMMC & 80.59\% & 7.94\% & 16.99\%\\
                \bottomrule
        \end{tabular}
        \label{tab:metrics}
\end{table}

\section{Discussion}

The accuracy for test data tended to be higher for the more complex models.
In particular, Deep MLP among the discriminative classifiers and SMMC among the generative classifiers achieved better performance than the other classifiers on average, thus indicating their suitability for EMG classification.
By contrast, LDA, which is widely used in EMG classification, demonstrated low accuracy for all datasets.
The strong constraint of sharing covariance among classes in the LDA is likely to adversely affect the classification of these datasets.

However, accuracy and confidence metrics are not necessarily positively (or negatively) correlated.
Discriminative classifiers, including Deep MLP, which shows high accuracy, tend to have worse ECE and MCE overall.
This can be observed from the reliability diagrams, indicating that neural network-based classifiers such as MLP and Deep MLP are overconfident.
Such overconfidence in highly nonlinear networks is consistent with phenomena reported in large image and natural language datasets~\cite{Guo2017-ml}.
By contrast, generative classifiers, in particular LDA and SMMC, showed better results in terms of ECE and MCE.
Further analysis is required to fully understand the underlying reasons. 
However, generative classifiers involve stronger assumptions regarding the data generative process; thus, if this assumption is consistent with the actual data distribution, then the confidence may approach the true predictive uncertainty.

Remarkably, SMMC demonstrated superior performance in terms of both accuracy and confidence.
In particular, ECE was less than 10\% for all datasets, indicating relatively few miscalibrations.
This might be because SMMC is constructed based on the generation characteristics of  EMG signals and can thus account for the uncertainty involved in EMG.
These results suggest that SMMC can effectively perform advanced tasks that utilize classifier confidence, such as the sequential learning, out-of-distribution detection, and active learning of EMG patterns.

\section{Conclusion}

In this paper, we conducted a visual and quantitative evaluation of the prediction confidence of classifiers for EMG pattern recognition.
We demonstrated that whereas complex discriminative classifiers exhibited high accuracy, their confidence deviated from the true probability.
Furthermore, we revealed that a scale mixture model-based generative classifier, SMMC, exhibited superior performance in terms of both accuracy and confidence.
This paper provides a solid foundation for future investigations pertaining to EMG pattern recognition and the use of confidence in classifiers.

A limitation of this study is that the classifiers and datasets employed were not exhaustive; thus, further enhancements are warranted.
Furthermore, neural network-based classifiers have been shown to improve confidence via the utilization of suitable calibration methods; hence, we plan to examine these effects.
Finally, the well-calibrated confidence output by the SMMC should be investigated.

\bibliographystyle{IEEEtran}
\bibliography{Ref_EMBC2023}

\addtolength{\textheight}{-12cm}   



\end{document}